\documentclass[twocolumn,aps,pra,showpacs,tightenlines]{revtex4-1}
\usepackage{amsmath}
\usepackage{amsfonts}
\usepackage{graphicx}
\usepackage{units}
\usepackage{color}
\usepackage[colorlinks,linkcolor=blue,anchorcolor=blue,citecolor=blue,urlcolor=blue]{hyperref}
\usepackage{multirow}
\begin{document}
\title{Retrieval of photon blockade effect in the dispersive Jaynes-Cummings model}

\author{Ya-Ting Guo}
\email{These authors contributed equally to this work.}
\affiliation{Key Laboratory of Low-Dimensional Quantum Structures and Quantum Control of
Ministry of Education, Key Laboratory for Matter Microstructure and Function of Hunan Province, Department of Physics and Synergetic Innovation Center for Quantum Effects and Applications, Hunan Normal University, Changsha 410081, China}
\author{Fen Zou}
\email{These authors contributed equally to this work.}
\affiliation{Key Laboratory of Low-Dimensional Quantum Structures and Quantum Control of
Ministry of Education, Key Laboratory for Matter Microstructure and Function of Hunan Province, Department of Physics and Synergetic Innovation Center for Quantum Effects and Applications, Hunan Normal University, Changsha 410081, China}
\author{Jin-Feng Huang}
\email{Corresponding author: jfhuang@hunnu.edu.cn}
\affiliation{Key Laboratory of Low-Dimensional Quantum Structures and Quantum Control of
Ministry of Education, Key Laboratory for Matter Microstructure and Function of Hunan Province, Department of Physics and Synergetic Innovation Center for Quantum Effects and Applications, Hunan Normal University, Changsha 410081, China}

\author{Jie-Qiao Liao}
\email{Corresponding author: jqliao@hunnu.edu.cn}
\affiliation{Key Laboratory of Low-Dimensional Quantum Structures and Quantum Control of
Ministry of Education, Key Laboratory for Matter Microstructure and Function of Hunan Province, Department of Physics and Synergetic Innovation Center for Quantum Effects and Applications, Hunan Normal University, Changsha 410081, China}

\date{\today}

\begin{abstract}
We propose a reliable scheme to recover the conventional photon blockade effect in the dispersive Jaynes-Cummings model, which describes a two-level atom coupled to a single-mode cavity field in the large-detuning regime. This is achieved by introducing a transversal driving to the atom and then photonic nonlinearity is obtained. The eigenenergy spectrum of the system is derived analytically and the photon blockade effect is confirmed by numerically calculating the photon-number distributions and equal-time second-order correlation function of the cavity field in the presence of system dissipations. We find that the conventional photon blockade effect can be recovered at proper atomic and cavity-field drivings. This work will provide a method to generate the conventional photon blockade effect in the dispersively coupled quantum optical systems.
\end{abstract}
\maketitle

\section{Introduction}
The photon blockade (PB) effect~\cite{Imamoglu1997}, as one of the physical methods for generation of single photons, has received considerable research interest in the past two decades. Depending on the underlying physical mechanisms, conventional~\cite{Imamoglu1997} and unconventional PBs~\cite{Liew2010,Bamba2011} have been proposed. Physically, the conventional PB effect is induced by the anharmonicity in the eigenenergy spectrum of the physical systems, while the unconventional PB effect is created by the quantum interference effect among different transition paths existing in physical systems.

By far, the conventional PB effect has been studied mostly in various nonlinear quantum optical systems, such as cavity quantum electrodynamical (QED) systems~\cite{Tian1992,Faraon2010,Ridolfo2012,Bajcsy2013,Miranowicz2014,Radulaski2017,Wang2017,Han2018,Trivedi2019,Hou2019,Zou2020PRA} and circuit QED systems~\cite{Liu2014}, which describe the interaction between fields and atoms. In addition, the conventional PB effect has been studied in coupled waveguide systems~\cite{Zheng2011,Huang2013} and various nonlinear bosonic-mode systems, such as Kerr-type nonlinear cavities~\cite{Ferretti2010,Liao2010,Miranowicz2013,Huang2018,Ghosh2019}, cavity optomechanical systems~\cite{Rabl2011,Liao2013,Liao2013JQ,Komar2013,Lv2013,Wang2015,Zhu2018,Zou2019}, and coupled nonlinear cavities~\cite{Majumdar2013,Zou2020,Deng2020}. In particular, the conventional PB effect has been experimentally demonstrated in coupled atom-field systems, including a single two-level atom coupled to an optical cavity~\cite{birnbaum2005,peyronel2012}, a quantum dot coupled to a photonic-crystal cavity~\cite{faraon2008,reinhard2012}, and a single superconducting artificial atom coupled to a transmission-line resonator~\cite{hoffman2011,lang2011}. In parallel, the unconventional PB effect has been studied theoretically~\cite{Liew2010,Bamba2011,Xu2013,Xu2014,Lemonde2014,Gerace2014,Zhang2014,Tang2015,Liu2016,Shen2015,Flayac2017,Zhou2016,Sarma2017,Ryou2018,Sarma2018,Li2019} and experimentally~\cite{Snijders2018,Vaneph2018} in various coupled quantum optical systems.

In coupled atom-cavity-field systems, both the resonant and off-resonant couplings have been studied and the optimal drivings for the PB effect have been found in these cases~\cite{birnbaum2005,Muller2015PRL,Muller2015PRX,Dory2017PRA,Liang2018}.
For example, the photon blockade effect has been experimentally demonstrated in the resonantly coupled atom-cavity-field system~\cite{birnbaum2005} and it has been found that there are two optimal drivings for photon blockade corresponding to the single-photon resonant transition. In addition, the influence of the detuning between the atom and the cavity field on the cavity-field statistics has recently been investigated~\cite{Muller2015PRL,Muller2015PRX,Dory2017PRA,Liang2018}. It has been found that there exist some optimal detunings corresponding to smaller values of the second-order correlation function than that in the resonant atom-field coupling case~\cite{Muller2015PRL}. It should be emphasized that these detuning-coupled atom-field systems do not enter the large-detuning regime, in which the system should be described by the dispersive JC model.

In the dispersive JC model, the conventional PB effect disappears because there is no photonic nonlinearity~\cite{footnote}. Therefore, an interesting question arising here is whether it is possible to retrieve the PB effect in the dispersive JC model using proper quantum manipulation. In this paper we propose an experimentally accessible method to recover the photonic nonlinearity in the system via introducing a transversal driving to the two-level atom. We obtain the analytical energy spectrum of the model, which gives the analytical result of the optimal driving frequency of the cavity field. We also get the steady state of the system by numerically solving the quantum master equation in the open-system case. We further calculate the photon-number distributions and the equal-time second-order correlation functions of the cavity field in the steady state. The results indicate that the PB effect can be recovered under proper atomic and cavity-field drivings. We also evaluate the validity of the driven dispersive JC model by presenting the detailed derivation of the effective Hamiltonian and checking the fidelity between the exact and approximate states, which are governed by the effective Hamiltonian and the approximate Hamiltonian, respectively. This work will provide an experimentally implementable method for the observation of the PB effect in dispersively coupled atom-field systems.

The rest of this paper is organized as follows. In Sec.~\ref{modelsec} we introduce the physical model and present the Hamiltonian. In Sec.~\ref{pbsec} we investigate the PB effect by calculating the photon-number distributions and the equal-time second-order correlation function of the cavity field with the method of the quantum master equation. In Sec.~\ref{discussion} we discuss the detailed derivation of the effective Hamiltonian. We also evaluate the validity of the approximate Hamiltonian by calculating the fidelity between the exact and approximate states. A discussion of the experimental implementation of this model is presented in Sec.~\ref{discussionexp}. Finally, we conclude this work with a summary in Sec.~\ref{conclusion}.
\begin{figure}[tbp]
\centering
\includegraphics[width=0.47\textwidth]{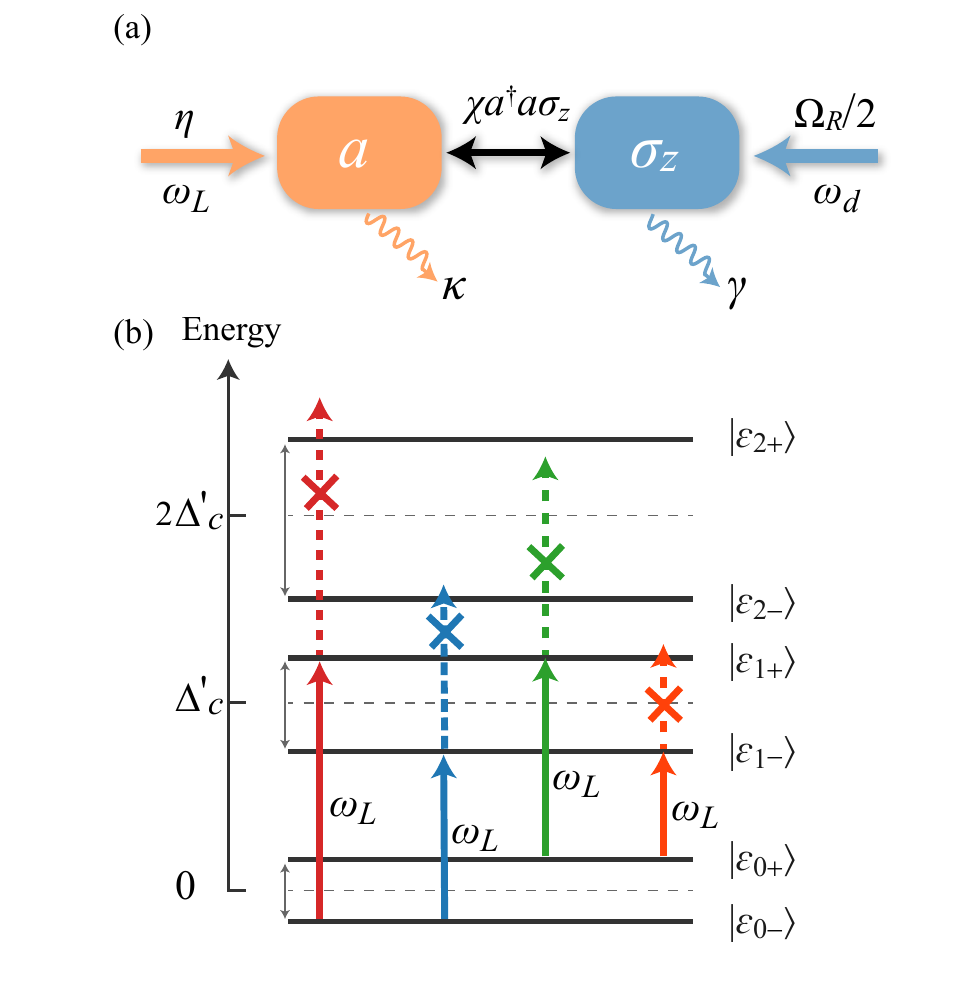}
\caption{(a) Schematic of the driven dispersive JC model, which is composed of a single-mode cavity field coupled to a two-level atom via the dispersive JC type of interaction. A transversal driving is applied to the atom to recover the optical nonlinearity in this system. (b) Diagram of the eigenenergy levels of the Hamiltonian $H^{(I)}_{\text{ddJC}}$ in the subspaces associated with zero, one, and two photons.}
\label{Fig1}
\end{figure}

\section{Model and Hamiltonian \label{modelsec}}

We consider a driven dispersive JC model consisting of a single-mode cavity field and a two-level atom, as shown in Fig.~\ref{Fig1}(a). Here the cavity field is coupled to the atom via a dispersive JC type of interaction, and a monochromatic field is applied to transversally drive the atom. The Hamiltonian of the driven dispersive JC model reads ($\hbar=1$)~\cite{Szombati2020}
\begin{align}
H_{\text{ddJC}}&=\omega_{c}a^{\dag}a+\frac{\omega_{0}+\chi}{2}\sigma_{z}+\chi a^{\dagger}a\sigma_{z} \nonumber \\
&\quad+\frac{\Omega_{R}}{2}(\sigma_{+}e^{-i\omega_{d}t}+\sigma_{-}e^{i\omega_{d}t}),\label{HdJC}
\end{align}
where $a$ ($a^{\dag}$) is the annihilation (creation) operator of the single-mode cavity field with resonance frequency $\omega_{c}$. The operators $\sigma_{z}=\sigma_{+}\sigma_{-}-\sigma_{-}\sigma_{+}$ and $\sigma_{x}=\sigma_{+}+\sigma_{-}$ are, respectively, the Pauli operators of the $z$ and $x$ directions for the two-level atom with bare transition frequency $\omega_{0}$ and energy shift $\chi$. Here $\sigma_{+}=\vert e\rangle\langle g\vert$ ($\sigma_{-}=\vert g\rangle\langle e\vert$) is the raising (lowering) operator defined based on the ground state $\vert g\rangle$ and excited state $\vert e\rangle$ of the atom. The third term in Eq.~(\ref{HdJC}) describes the dispersive JC type of interaction between the cavity mode and the atom with the coupling strength $\chi$. The last term in Eq.~(\ref{HdJC}) denotes a monochromatic driving to the atom, with $\Omega_{R}/2$ and $\omega_{d}$ the driving strength and driving frequency, respectively. Here the motivation for introducing the transversal driving to the atom is to recover the photonic nonlinearity, which provides the physical mechanism for generation of the conventional PB effect. Note that the present driven dispersive JC model was recently implemented in a circuit QED system, in which a superconducting resonator was coupled dispersively to a superconducting transmon qubit~\cite{Szombati2020}. We will present a detailed derivation of the driven dispersive JC Hamiltonian and evaluate the quality of the approximate Hamiltonian in Sec.~\ref{discussion}.

In a rotating frame with respect to $\omega_{d}(a^{\dagger}a+\sigma_{z}/2)$, the Hamiltonian $H_{\text{ddJC}}$ becomes
\begin{equation}
H^{(I)}_{\text{ddJC}}=\Delta_{c}^{\prime}a^{\dag}a+\frac{\Delta_{0}}{2}\sigma_{z}+\chi a^{\dagger}a\sigma_{z}+\frac{\Omega_{R}}{2}\sigma_{x},\label{HdJCI}
\end{equation}
with $\Delta_{c}^{\prime}=\omega_{c}-\omega_{d}$ and $\Delta_{0}=\omega_{0}+\chi-\omega_{d}=\Delta_{0}^{\prime}+\chi$, where $\Delta_{c}^{\prime}$ ($\Delta_{0}^{\prime}=\omega_{0}-\omega_{d}$) is the driving detuning of the cavity-field resonance (atomic bare transition) frequency $\omega_{c}$ ($\omega_{0}$) with respect to the atomic-driving frequency $\omega_{d}$. In the following, we consider the case in which the cavity field is coherently driven by a monochromatic field with driving frequency $\omega_{L}$ and driving strength $\eta$. In this case, the total Hamiltonian of the whole system in the Schr\"{o}dinger picture can be expressed as $H_{\text{tot}}=H_{\text{ddJC}}+\eta(a^{\dag}e^{-i\omega_{L}t}+ae^{i\omega_{L}t})$. In a rotating frame defined by the unitary operator $\exp[-i\omega_{d}t(a^{\dagger}a+\sigma_{z}/2)-i\omega_{L}ta^{\dagger}a]$, the Hamiltonian $H_{\text{tot}}$ becomes
\begin{equation}
H_{I}=\Delta_{c}a^{\dag}a+\frac{\Delta_{0}}{2}\sigma_{z}+\frac{\Omega_{R}}{2}\sigma_{x}+\chi a^{\dag}a\sigma_{z}+\eta(a^{\dag}+a),\label{hams}
\end{equation}
where $\Delta_{c}=\Delta_{c}^{\prime}-\omega_{L}$ is the difference between the driving detuning $\Delta_{c}^{\prime}$ and the cavity-field driving frequency $\omega_{L}$.

To analyze the PB effect in this system, we need to know the eigensystem of the driven dispersive JC Hamiltonian $H^{(I)}_{\text{ddJC}}$, which takes a time-independent
form in the rotating frame. The eigensystem of the Hamiltonian $H^{(I)}_{\text{ddJC}}$ can be obtained as
\begin{equation}
H^{(I)}_{\text{ddJC}}\vert\varepsilon_{m,\pm}\rangle=E_{m,\pm}\vert\varepsilon_{m,\pm}\rangle,
\end{equation}
where the eigenvalues are given by
\begin{equation}
E_{m,\pm}=m\Delta_{c}^{\prime}\pm\frac{1}{2}\sqrt{(\Delta_{0}+2m\chi)^{2}+\Omega_{R}^{2}}\label{value}
\end{equation}
and the corresponding eigenstates are defined by $\vert \varepsilon_{m,\pm}\rangle=\vert m\rangle_{a}\vert\pm(m)\rangle$. Here $\vert m\rangle_{a}$ ($m=0,1,2,\ldots$) are the photon number states and the photon-number-dependent atomic states $\vert\pm(m)\rangle$ are given by
\begin{subequations}
\begin{align}
\vert+(m)\rangle=&\cos\theta_{m}\vert e\rangle+\sin\theta_{m}\vert g\rangle,\\
\vert-(m)\rangle=&-\sin\theta_{m}\vert e\rangle+\cos\theta_{m}\vert g\rangle,
\end{align}
\end{subequations}
where the mixing angle $\theta_{m}$ is defined by $\tan(2\theta_{m})=\Omega_{R}/(\Delta_{0}+2m\chi)$.

It can be seen from Eq.~(\ref{value}) that the eigenenergy of the system is a nonlinear function of the photon number $m$. This nonlinearity in the eigenenergy spectrum is the physical origin of the PB effect in this system. However, the anharmonicity will become weak in both the weak- and strong-driving cases. This point can be seen from the fact that the eigenvalues are reduced to $E_{m,\pm}\approx m\Delta_{c}^{\prime}\pm(\Delta_{0}+2m\chi)/2$ and $E_{m,\pm}\approx m\Delta_{c}^{\prime}\pm\Omega_{R}/2$ in the cases of $\Omega_{R}\ll\Delta_{0}+2m\chi$ and $\Omega_{R}\gg\Delta_{0}+2m\chi$, respectively. In these two cases, the energy spectrum of the photonic part is approximately harmonic. In Fig.~\ref{Fig1}(b) we schematically show the eigenenergies $E_{m,\pm}$ of the system in the few-photon subspaces ($m=0,1,2$). It can be seen that the energy spectrum of the driven dispersive JC model in the rotating frame is anharmonic, which motivates us to study the PB effect in this system.

\section{Photon blockade effect \label{pbsec}}

In this section we study the PB effect in this system by numerically calculating the photon-number distributions and the equal-time second-order correlation function of the cavity-field mode. In particular, we obtain the steady state of the system and the field statistics by numerically solving the quantum master equation in the open-system case.
\begin{figure}[tbp]
\centering
\includegraphics[width=0.47\textwidth]{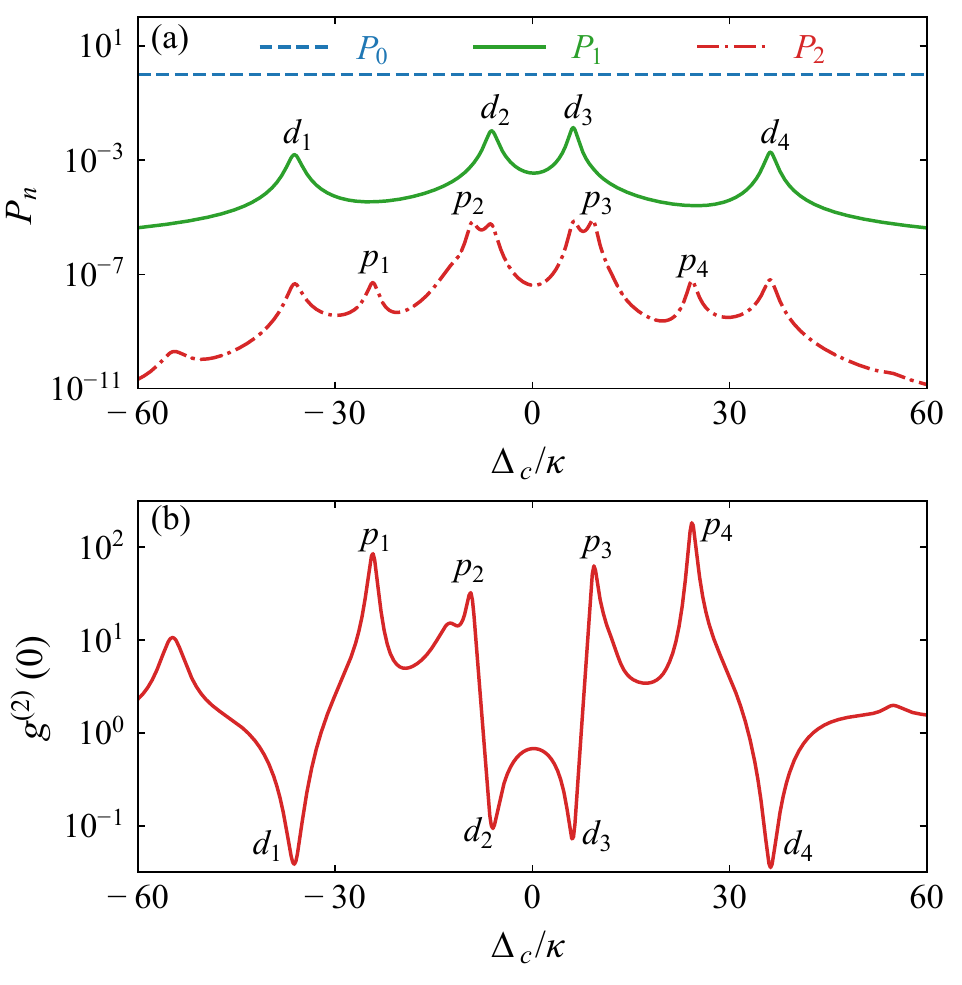}
\caption{(a) Photon-number distributions $P_{n=0,1,2}$ as functions of the scaled driving detuning $\Delta_{c}/\kappa$. (b) Equal-time second-order correlation function $g^{(2)}(0)$ as a function of the scaled driving detuning $\Delta_{c}/\kappa$. The other parameters are $\gamma/\kappa=0.5$, $\chi/\kappa=15$, $\Omega_{R}/\chi=2$, $\Delta_{0}=0$, and $\eta/\kappa=0.1$.}
\label{Fig2}
\end{figure}

 In a realistic situation, the physical system will inevitably interact with its environment. In this work we assume that the cavity field and the two-level atom are connected with two individual vacuum baths. Then the dynamics of the system is governed by the quantum master equation
\begin{equation}
\dot{\rho}=i[\rho,H_{I}]+\mathcal{D}_{a}[\rho]+\mathcal{D}_{\sigma}[\rho],
\label{qme}
\end{equation}
where $\rho$ is the density matrix of the system, $H_{I}$ is given by Eq.~(\ref{hams}), and the Lindblad superoperators are given by~\cite{Scully}
\begin{subequations}
\begin{align}
\mathcal{D}_{a}[\rho]&=\frac{\kappa}{2}(2a\rho a^{\dag}-a^{\dag}a\rho-\rho a^{\dag}a),\\
\mathcal{D}_{\sigma}[\rho]&=\frac{\gamma}{2}(2\sigma_{-}\rho \sigma_{+}-\sigma_{+}\sigma_{-}\rho-\rho \sigma_{+}\sigma_{-}),
\end{align}
\end{subequations}
with $\kappa$ ($\gamma$) the decay rate of the cavity-field mode (atom). By numerically solving Eq.~(\ref{qme}), the steady-state density operator $\rho_{\text{ss}}$ of the system can be obtained~\cite{johansson2012QuTiP,johansson2013QuTiP}. Then the photon-number distributions $P_{n}=\text{Tr}[\vert n\rangle_{a}\,_{a}\langle n\vert \rho_{\text{ss}}]$ and the equal-time second-order correlation function $g^{(2)}(0)=\text{Tr}(a^{\dagger2}a^{2}\rho_{\text{ss}})/[\text{Tr}(a^{\dagger}a\rho_{\text{ss}})]^2$ can be obtained accordingly.

To study the statistical properties of the cavity field, we plot the photon-number distributions $P_{n=0,1,2}$ as functions of the scaled driving detuning $\Delta_{c}/\kappa$ in Fig.~\ref{Fig2}(a). We can see that the photon-number distributions satisfy $P_{0}\approx1$ and $P_{0}\gg P_{1}\gg P_{2}$ in the weak-driving case. In addition, it can be seen that there are some peaks in the curves of $P_{1}$ and $P_{2}$. By analyzing the energy spectrum of the system [see Fig.~\ref{Fig1}(b)], we find that the locations of the four peaks $d_{n=1,2,3,4}$ in $P_{1}$ correspond to the single-photon resonant transitions $\vert \varepsilon_{0,+}\rangle\rightarrow\vert \varepsilon_{1,\pm}\rangle$ and $\vert \varepsilon_{0,-}\rangle\rightarrow\vert \varepsilon_{1,\pm}\rangle$, with the resonance conditions $\Delta_{c}=\Delta^{1,\pm}_{0,+}=[\mp\sqrt{(\Delta_{0}+2\chi)^{2}+\Omega_{R}^{2}}+\sqrt{\Delta_{0}^2+\Omega_{R}^{2}}]/2$ and $\Delta_{c}=\Delta^{1,\pm}_{0,-}=[\mp\sqrt{(\Delta_{0}+2\chi)^{2}+\Omega_{R}^{2}}-\sqrt{\Delta_{0}^2+\Omega_{R}^{2}}]/2$, respectively. Moreover, the locations of the four main peaks $p_{n=1,2,3,4}$ in $P_{2}$ are associated with the two-photon resonant transitions $\vert\varepsilon_{0,+}\rangle\rightarrow\vert \varepsilon_{2,\pm}\rangle$ and $\vert \varepsilon_{0,-}\rangle\rightarrow\vert \varepsilon_{2,\pm}\rangle$; the corresponding resonance conditions are given by $\Delta_{c}=\Delta^{2,\pm}_{0,+}=[\mp\sqrt{(\Delta_{0}+4\chi)^{2}+\Omega_{R}^{2}}+\sqrt{\Delta_{0}^2+\Omega_{R}^{2}}]/4$ and $\Delta_{c}=\Delta^{2,\pm}_{0,-}=[\mp\sqrt{(\Delta_{0}+4\chi)^{2}+\Omega_{R}^{2}}-\sqrt{\Delta_{0}^2+\Omega_{R}^{2}}]/4$, respectively. In addition, the other two peaks in $P_{2}$ are induced by the single-photon resonant transitions.

Usually, the PB effect will sensitively depend on the frequency of the cavity-field driving. This is because this frequency determines the resonance of the single-photon physical transitions. To find the optimal driving frequency of the PB effect, in Fig.~\ref{Fig2}(b) we plot the equal-time second-order correlation function $g^{(2)}(0)$ as a function of the scaled driving detuning $\Delta_{c}/\kappa$. It can be observed that the locations of these dips $d_{n}$ and peaks $p_{n}$ in the curve of $g^{(2)}(0)$ correspond to the single- and two-photon resonant transitions, respectively. In particular, we find that the optimal PB effect [the correlation function $g^{(2)}(0)\ll1$] takes place at single-photon resonant transitions $\Delta_{c}=\Delta^{1,\pm}_{0,+}$ and $\Delta_{c}=\Delta^{1,\pm}_{0,-}$ (the location of $d_{n}$).
\begin{figure}[tbp]
\centering
\includegraphics[width=0.47\textwidth]{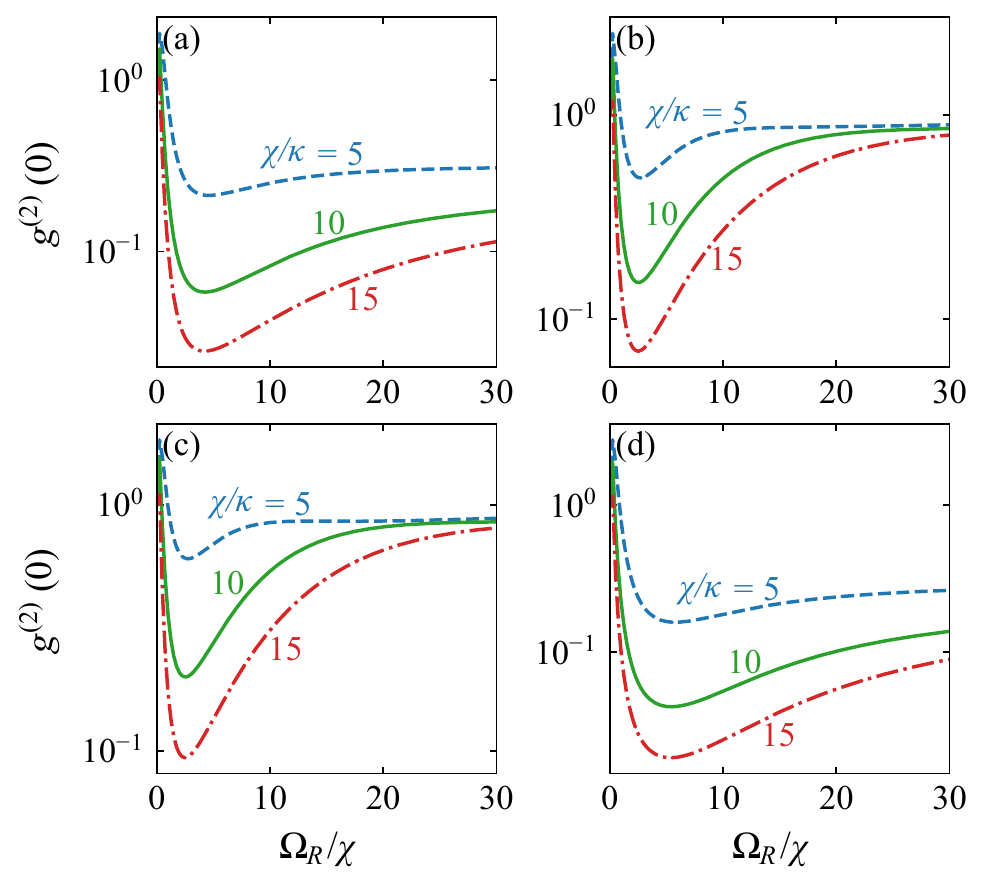}
\caption{Equal-time second-order correlation function $g^{(2)}(0)$ as a function of the ratio $\Omega_{R}/\chi$ of the atomic-driving strength $\Omega_{R}$ over the dispersive coupling strength $\chi$ at $\chi/\kappa=5$, $10$, and $15$ for four panels correspond to four single-photon resonant transition cases: (a) $\Delta_{c}=\Delta^{1,+}_{0,-}$, (b) $\Delta_{c}=\Delta^{1,-}_{0,-}$, (c) $\Delta_{c}=\Delta^{1,+}_{0,+}$, and (d) $\Delta_{c}=\Delta^{1,-}_{0,+}$. The other parameters are $\gamma/\kappa=0.5$, $\Delta_{0}=0$, and $\eta/\kappa=0.1$.}
\label{Fig3}
\end{figure}

As analyzed in Sec.~\ref{modelsec}, the photonic nonlinearity will disappear in the weak- and strong-atomic-driving cases. Therefore, it is an interesting topic to study the influence of the atomic-driving strength on the PB effect. In Fig.~\ref{Fig3} the equal-time second-order correlation function $g^{(2)}(0)$ is plotted as a function of the ratio $\Omega_{R}/\chi$ of the atomic-driving strength $\Omega_{R}$ over the dispersive coupling strength $\chi$ at different single-photon resonant transitions $\Delta_{c}=\Delta^{1,\pm}_{0,+}$ and $\Delta_{c}=\Delta^{1,\pm}_{0,-}$. Here the blue dashed curves, green solid curves, and red dash-dotted curves correspond to the scaled dispersive coupling strengths $\chi/\kappa=5$, $10$, and $15$, respectively. We can see that the envelope of the correlation functions is lower for a larger value of the scaled dispersive coupling strength $\chi/\kappa$. This means that the PB effect is stronger for a larger dispersive JC coupling strength. In addition, for a given $\chi$, as the atomic-driving strength $\Omega_{R}/\chi$ increases, the values of the correlation functions first decrease and then increase. This indicates that there is an optimal transversal driving strength of the atom. Physically, around the optimal atomic driving, the photon nonlinearity in the eigenenergy spectrum is strong. Here we can see that the optimal value of $\Omega_{R}/\chi$ is around $2$. This is because the photonic nonlinearity becomes important when $\Omega_{d}$ is almost large as $2m\chi$, where the contributing $m$ is $1$ in the photon blockade regime. In the weak- and strong-atom-driving cases, the PB effect disappears gradually. As shown in Fig.~\ref{Fig3}, the correlation function $g^{(2)}(0)$ approaches $1$ in the limit cases $\Omega_{R}/\chi\rightarrow0$ and $\Omega_{R}/\chi\gg2m$.
\begin{figure}[tbp]
\centering
\includegraphics[width=0.47\textwidth]{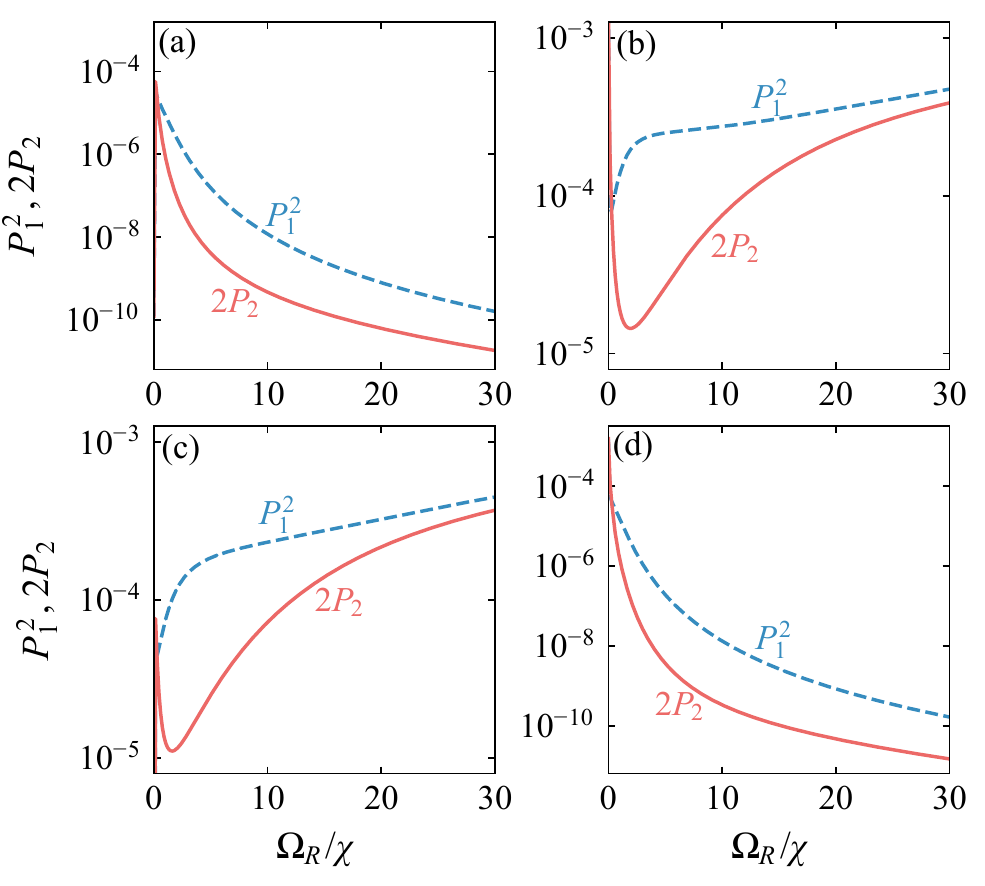}
\caption{Variables $2P_{2}$ and $P^{2}_{1}$ versus the ratio $\Omega_{R}/\chi$ for four panels correspond to four cases of single-photon resonant transition: (a) $\Delta_{c}=\Delta^{1,+}_{0,-}$, (b) $\Delta_{c}=\Delta^{1,-}_{0,-}$, (c) $\Delta_{c}=\Delta^{1,+}_{0,+}$, and (d) $\Delta_{c}=\Delta^{1,-}_{0,+}$. The other parameters are $\gamma/\kappa=0.5$, $\Delta_{0}=0$, $\chi/\kappa=15$, and $\eta/\kappa=0.1$.}
\label{Fig4}
\end{figure}

The optimal atomic-driving phenomenon can also be quantitatively explained by analyzing the photon-number distributions. In the weak-driving regime, the involved photon number is small and the second-order correlation function can be approximately expressed with the single- and two-photon distributions as $g^{(2)}(0)\approx 2P_{2}/P^{2}_{1}$. In Fig.~\ref{Fig4} we plot $2P_{2}$ and $P^{2}_{1}$ [the numerator and denominator of $g^{(2)}(0)$] as functions of the ratio $\Omega_{R}/\chi$ at different single-photon resonant transitions $\Delta_{c}=\Delta^{1,\pm}_{0,+}$ and $\Delta_{c}=\Delta^{1,\pm}_{0,-}$, corresponding to the cases ($\chi/\kappa=15$) in the four panels of Fig.~\ref{Fig3}. It can be seen from Figs.~\ref{Fig4}(b) and~\ref{Fig4}(c) that $P^{2}_{1}$ is a monotonically increasing function of the ratio $\Omega_{R}/\chi$, but $2P_{2}$ first decreases and then increases with the increase of $\Omega_{R}/\chi$. As a result, the second-order correlation function $g^{(2)}(0)$ has a similar curve monotonicity as $2P_{2}$, corresponding to the dash-dotted curves in Figs.~\ref{Fig3}(b) and~\ref{Fig3}(c). In Figs.~\ref{Fig4}(a) and~\ref{Fig4}(d), both $2P_{2}$ and $P^{2}_{1}$ are monotonically decreasing functions of the ratio $\Omega_{R}/\chi$. Since the curve of the numerator $2P_{2}$ has a larger curvature than that of the denominator $P^{2}_{1}$, the value of the correlation function $g^{(2)}(0)$ will experience a change of the monotonicity. With the increase of $\Omega_{R}/\chi$, $2P_{2}$ first decreases faster than $P^{2}_{1}$ and then decreases slower than $P^{2}_{1}$. Consequently, $g^{(2)}(0)$ first decreases and then increases with the increase of the ratio $\Omega_{R}/\chi$.

To clearly explain the physical reason for the optimal atomic-driving effect, it will be helpful to further analyze the dependence of the photon-number distributions $P_{1}$ and $P_{2}$ on the driving strength $\Omega_{R}$. In the weak cavity-field driving regime, the system can be reduced to a six-level system by restricting the system to the zero-, one-, and two-photon subspaces, as shown in Fig.~\ref{Fig1}(b). Based on the relations $\vert \varepsilon_{m,\pm}\rangle=\vert m\rangle_{a}\vert\pm(m)\rangle$, the photon-number distributions $P_{m}$ can be obtained as $P_{m}\equiv\text{Tr}[\vert m\rangle_{a}\,_{a}\langle m\vert \rho_{\text{ss}}]=P_{m,+}+P_{m,-}$, where $P_{m,\pm}=\text{Tr}[\vert \varepsilon_{m,\pm}\rangle\langle \varepsilon_{m,\pm}\vert \rho_{\text{ss}}]$ are the populations of the eigenstates. As a result, the photon-number distributions can be obtained by calculating the population of these six eigenstates $\vert \varepsilon_{m,\pm}\rangle$ for $n=0$, $1$, and $2$. For this six-level system, we cannot obtain the analytical result of the steady-state populations, and hence it is hard to know the dependence of the photon-number distributions $P_{m}$ on the atomic-driving strength $\Omega_{R}$. However, we roughly estimated the detuning of the second-photon transitions and found that the populations $P_{2,\pm}$ of the two-photon eigenstates $\vert \varepsilon_{2,\pm}\rangle$ and the corresponding detunings have an inverse dependence on the atomic-driving strength $\Omega_{R}$. This phenomenon can qualitatively explain the optimal atomic-driving effect, because a large detuning will suppress the quantum transition induced by the absorbtion of the second photon.

\begin{figure}[tbp]
\centering
\includegraphics[width=0.47\textwidth]{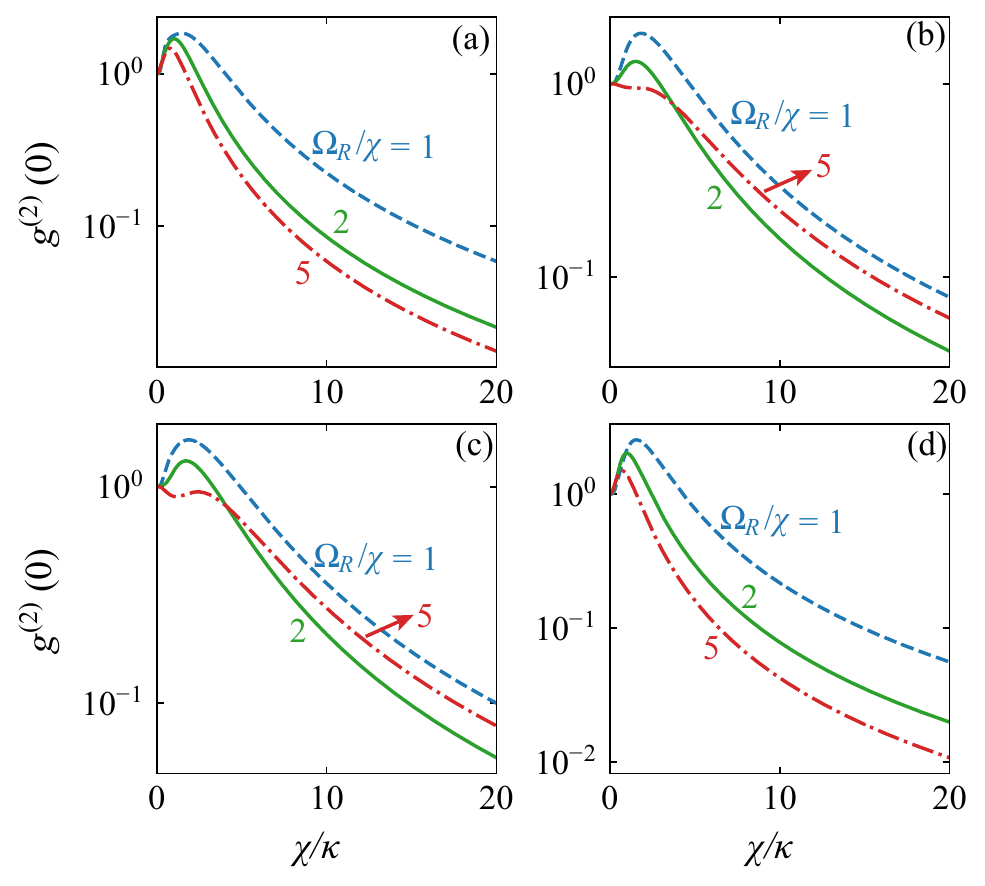}
\caption{Equal-time second-order correlation function $g^{(2)}(0)$ as a function of the scaled dispersive coupling strength $\chi/\kappa$ at $\Omega_{R}/\chi=1$, $2$, and $5$ for four panels correspond to four single-photon resonant transition cases: (a) $\Delta_{c}=\Delta^{1,+}_{0,-}$, (b) $\Delta_{c}=\Delta^{1,-}_{0,-}$, (c) $\Delta_{c}=\Delta^{1,+}_{0,+}$, and (d) $\Delta_{c}=\Delta^{1,-}_{0,+}$. The other parameters are $\gamma/\kappa=0.5$, $\Delta_{0}=0$, and $\eta/\kappa=0.1$.}
\label{Fig5}
\end{figure}

We also analyze the dependence of the PB effect on the dispersive JC coupling strength $\chi$. In Fig.~\ref{Fig5} we plot the equal-time second-order correlation function $g^{(2)}(0)$ as a function of the ratio $\chi/\kappa$ at different single-photon resonant transitions $\Delta_{c}=\Delta^{1,\pm}_{0,+}$ and $\Delta_{c}=\Delta^{1,\pm}_{0,-}$. Here the blue dashed curves, green solid curves, and red dash-dotted curves correspond to the ratios $\Omega_{R}/\chi=1$, $2$, and $5$, respectively. It can be seen that the values of the correlation function $g^{(2)}(0)$ first increase and then decrease with the increase of the ratio $\chi/\kappa$. The reason is that the larger the scaled dispersive coupling strength $\chi/\kappa$, the stronger the nonlinearity of the driven dispersive JC model. For generation of a strong PB effect, the system should work in the strong dispersive JC coupling regime $\chi/\kappa\gg 1$~\cite{Schuster2007Resolving}, which is a stronger parameter requirement than the strong-coupling condition $g\gg \kappa$ because of the relations $\chi=g^{2}/(\omega_{0}-\omega_{c})$, $\chi\gg\kappa$, and the large-detuning condition $|\omega_{0}-\omega_{c}|\gg g$. In addition, it can be seen from Figs.~\ref{Fig5}(b) and~\ref{Fig5}(c) that the values of the correlation functions at the atomic-driving strength $\Omega_{R}/\chi=5$ are larger than that at $\Omega_{R}/\chi=2$ in the cases of $\Delta_{c}=\Delta^{1,+}_{0,+}$ and $\Delta_{c}=\Delta^{1,-}_{0,-}$. This feature is a consequence of the nonmonotonicity of the dependence of the correlation function $g^{(2)}(0)$ on the driving strength $\Omega_{R}$.

\section{Validity of the driven dispersive JC Hamiltonian\label{discussion}}
In this section we discuss the validity of the driven dispersive JC Hamiltonian~(\ref{HdJCI}). Concretely, we present a detailed derivation of the effective Hamiltonian of the driven JC system working in the large-detuning regime. We also analyze the difference between the derived effective Hamiltonian and the driven dispersive JC Hamiltonian~(\ref{HdJCI}) by calculating the fidelity between the two states, which are governed by the derived Hamiltonian and the Hamiltonian~(\ref{HdJCI}), respectively.

\subsection{Derivation of the driven dispersive JC Hamiltonian}
Here we present a detailed derivation of the driven dispersive JC Hamiltonian based on the driven JC model, in which the atom is largely detuned coupled with the cavity field and the atom is driven by a monochromatic field. The Hamiltonian of the driven JC model reads
\begin{align}
H_{\text{dJC}}&=\omega_{c}a^{\dagger}a+\frac{\omega_{0}}{2}\sigma_{z}+g(a^{\dagger}\sigma_{-}+\sigma_{+}a)\nonumber\\
&\quad+\frac{\Omega_{R}}{2}(\sigma_{+}e^{-i\omega_{d}t}+\sigma_{-}e^{i\omega_{d}t}),\label{HJC}
\end{align}
where $g$ is the coupling strength between the cavity field and the atom, the operators and other parameters have been defined in Eq.~(\ref{HdJC}).

In a rotating frame with respect to $\omega_{d}(a^{\dagger}a+\sigma_{z}/2)$, the Hamiltonian $H_{\text{dJC}}$ can be expressed as
\begin{equation}
H_{\text{dJC}}^{(I)}=H_{0}+H_{I}+H_{d},
\end{equation}%
with
\begin{subequations}
\begin{align}
H_{0}&=\Delta_{c}^{\prime}a^{\dagger}a+\frac{\Delta_{0}^{\prime}}{2}\sigma_{z},\\
H_{I}&=g(a^{\dagger}\sigma_{-}+\sigma_{+}a),\\
H_{d}&=\frac{\Omega_{R}}{2}\sigma_{x},
\end{align}
\end{subequations}%
where $\Delta_{0}^{\prime}=\omega_{0}-\omega_{d}$ and $\Delta_{c}^{\prime}=\omega_{c}-\omega_{d}$ have been introduced in Eq.~(\ref{HdJCI}). In the large-detuning case $\vert\Delta\vert=\vert\Delta_{0}^{\prime}-\Delta_{c}^{\prime}\vert\gg g\sqrt{n+1}$, with $n$ the maximally dominated photon number involved in the cavity field, the Hamiltonian of the driven dispersive JC model can be obtained by using the Fr\"{o}hlich-Nakajima transformation~\cite{Frohlich1950Theory,Nakajima1955Perturbation}. To this end, we introduce an anti-Hermite operator $S$, which is determined by the equation $H_{I}+[H_{0},S]=0$. Then the anti-Hermite operator $S$ can be derived as
\begin{equation}
S=\frac{g}{\Delta}(a^{\dagger}\sigma_{-}-\sigma_{+}a).
\end{equation}
Up to the second order of the ratio $g/\Delta$, the effective Hamiltonian describing the system can be approximately obtained as
\begin{align}
H_{\text{eff}}&=e^{-S}H_{\text{dJC}}^{(I)}e^{S}  \nonumber \\
&\approx\Delta_{c}^{\prime}a^{\dagger}a+\frac{\Delta_{0}^{\prime}}{2}\sigma_{z}+\chi\left[\sigma_{z}a^{\dagger}a+\frac{1}{2}(\sigma_{z}+I)\right]+\frac{\Omega_{R}}{2}\sigma_{x}\nonumber\\
&\quad+\frac{\Omega_{R}}{2}\frac{g}{\Delta}\sigma_{z}(a^{\dagger}+a),
\end{align}
where we introduced the dispersive JC coupling strength $\chi=g^{2}/\Delta$.

To maintain consistency between the effective Hamiltonian $H_{\text{eff}}$ and the Hamiltonian given by Eq.~(\ref{HdJCI}), we introduce the variables $\Delta_{0}=\Delta_{0}^{\prime}+\chi$; then the effective Hamiltonian becomes
\begin{align}
H_{\text{eff}}&=\Delta_{c}^{\prime}a^{\dagger}a+\frac{\Delta_{0}}{2}\sigma_{z}+\chi\sigma_{z}a^{\dagger}a+\frac{\Omega_{R}}{2}\sigma_{x}\nonumber\\
&\quad+\frac{\Omega_{R}}{2}\frac{g}{\Delta}\sigma_{z}(a^{\dagger}+a),\label{Heff}
\end{align}
where we discarded the constant term $\chi/2$. The last term in Eq.~(\ref{Heff}) can be regarded as a conditional cavity-field driving term depending on the states of the two-level atom. Under the parameter condition of $\Omega_{R}g/2\vert\Delta\vert\ll\vert\Delta_{c}^{\prime}+\chi\vert$, the last term in Eq.~(\ref{Heff}) can be approximately neglected then the effective Hamiltonian~(\ref{Heff}) is reduced to the driven dispersive JC Hamiltonian given by Eq.~(\ref{HdJCI}).

\subsection{Evaluation of the validity of the driven dispersive JC model}

To quantitatively evaluate the validity of the driven dispersive JC Hamiltonian~(\ref{HdJCI}), we check the fidelity $F(t)=\vert\langle\psi(t)|\varphi(t)\rangle\vert^{2}$ between the state $\vert\psi(t)\rangle$ governed by the effective Hamiltonian~(\ref{Heff}) and the state $\vert\varphi(t)\rangle$ governed by the dispersive JC Hamiltonian~(\ref{HdJCI}). Below we derive the expression of the two states by solving the equations of motion for these probability amplitudes in the two cases.

Corresponding to the effective Hamiltonian~(\ref{Heff}), the state of this system at time $t$ is defined as
\begin{equation}
\vert\psi(t)\rangle=\sum^{\infty}_{n=0}[A_{n}(t)\vert e,n\rangle+B_{n}(t)\vert g,n\rangle],
\end{equation}
where $A_{n}(t)$ and $B_{n}(t)$ are the probability amplitudes. According to the Schr\"{o}dinger equation, the equations of motion for these amplitudes are obtained as
\begin{subequations}
\label{eqHeff}
\begin{align}
\dot{A}_{n}=&-i\left(n\Delta_{c}^{\prime}+\Delta_{0}/2+n\chi\right)A_{n}-i\frac{\Omega_{R}}{2}B_{n}\nonumber\\
&-i\frac{\Omega_{R}}{2}\frac{g}{\Delta}(\sqrt{n}A_{n-1}+\sqrt{n+1}A_{n+1}),\\
\dot{B}_{n}=&-i\left(n\Delta_{c}^{\prime}-\Delta_{0}/2-n\chi\right)B_{n}-i\frac{\Omega_{R}}{2}A_{n}\nonumber\\
&+i\frac{\Omega_{R}}{2}\frac{g}{\Delta}(\sqrt{n}B_{n-1}+\sqrt{n+1}B_{n+1}).
\end{align}
\end{subequations}

For the dispersive JC Hamiltonian~(\ref{HdJCI}), we assume that the state of the system takes the form
\begin{equation}
\vert\varphi(t)\rangle=\sum^{\infty}_{n=0}[a_{n}(t)\vert e,n\rangle+b_{n}(t)\vert g,n\rangle],
\end{equation}
with the probability amplitudes $a_{n}(t)$ and $b_{n}(t)$. The evolution of these probability amplitudes is determined by the equations
\begin{subequations}
\label{eqHdJCI}
\begin{align}
\dot{a}_{n}=&-i\left(n\Delta_{c}^{\prime}+\Delta_{0}/2+n\chi\right)a_{n}-i\frac{\Omega_{R}}{2}b_{n},\\
\dot{b}_{n}=&-i\left(n\Delta_{c}^{\prime}-\Delta_{0}/2-n\chi\right)b_{n}-i\frac{\Omega_{R}}{2}a_{n}.
\end{align}
\end{subequations}
By solving Eqs.~(\ref{eqHeff}) and~(\ref{eqHdJCI}) under a given initial state, we can obtain the states $\vert \psi(t)\rangle$ and $\vert\varphi(t)\rangle$ accordingly. Then the fidelity can be obtained as
\begin{eqnarray}
F(t)&=& \vert\langle\psi(t)|\varphi(t)\rangle\vert^{2}\nonumber\\
&=&\left\vert \sum^{\infty}_{n=0} [A^{*}_{n}(t)a_{n}(t)+B^{*}_{n}(t)b_{n}(t)] \right\vert^{2}.
\end{eqnarray}
Note that in a realistic calculation, we need to truncate the dimension of the Hilbert space of the cavity field up to a finite number, which is determined by the photon-number distribution in the cavity under a given initial state.
\begin{figure}[tbp]
\centering
\includegraphics[width=0.47\textwidth]{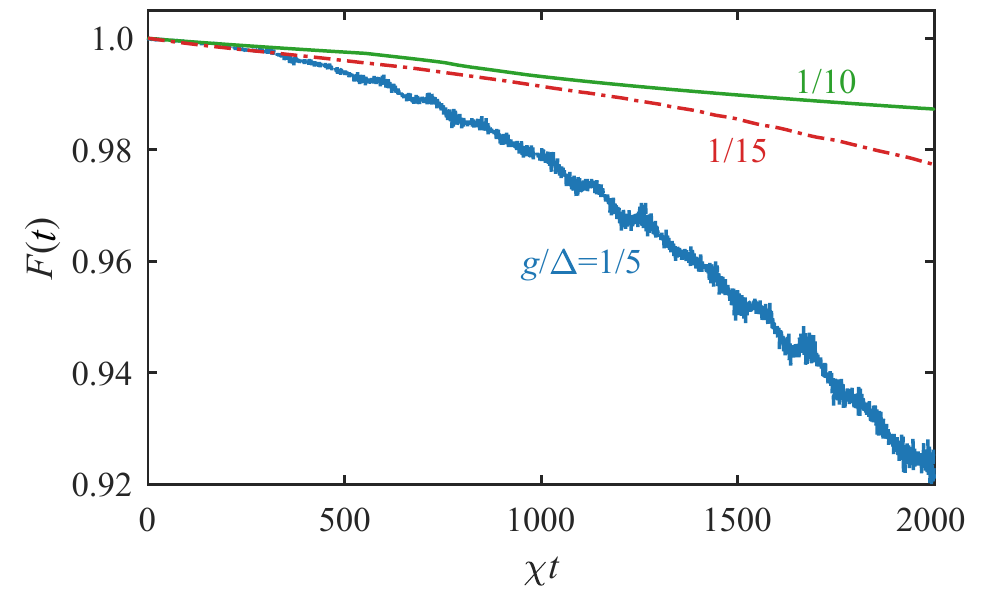}
\caption{Fidelity $F(t)$ as a function of the scaled time $\chi t$ at different values of the ratio $g/\Delta$. The other parameters are $\Delta_{0}=0$, $\Delta_{c}^{\prime}=\Delta_{0}-\chi-\Delta$, $\Delta_{c}^{\prime}/\chi=-1-(\Delta/g)^{2}$, and $\Omega_{R}/\chi=2$.}
\label{Fig6}
\end{figure}

To confirm the parameter conditions of the driven dispersive JC model, in Fig.~\ref{Fig6} we plot the fidelity $F(t)$ as a function of the evolution time $\chi t$ at different values of $g/\Delta$. Here we choose the initial state of the system as $\vert \psi(0)\rangle=\vert\alpha\rangle(\vert g\rangle+\vert e\rangle)/\sqrt{2}$, with $\vert\alpha\rangle$ the coherent state ($\alpha=1$). It can be seen that the fidelity is high for a large ratio of $\vert\Delta\vert/g$. This means that, in the large-detuning regime, the present physical system can be well described by the driven dispersive JC model.

\section{Experimental implementation of this scheme \label{discussionexp}}

In this section we discuss the experimental implementation of the present scheme. It can be seen from the Hamiltonian in Eq.~(\ref{hams}) that, to implement the present physical scheme, the candidate physical setups should be able to implement the dispersive JC interaction, the transversal atom driving, and the cavity-field driving. In particular, to observe the photon blockade effect in the dispersive JC model, the system should work in the strong dispersive JC interaction regime, in which the photon-number dependent qubit energy shift can be resolved. Generally speaking, both the atom driving and the cavity-field driving are experimentally accessible in most physical systems. Therefore, the key point for experimental implementation of this scheme should be the realization of the strong dispersive JC interaction.

Currently, the strong dispersive JC interaction has been experimentally implemented in a circuit-QED system~\cite{Szombati2020,Schuster2007Resolving}. Below we discuss the parameter analysis of the dispersive JC interaction in a circuit QED system consisting of a superconducting charge qubit and a coplanar superconducting transmission-line resonator~\cite{Schuster2007Resolving}. In this system, the resonant frequency of the coplanar cavity is $\omega_{r}/2\pi=5.7$ GHz and the decay rate of the cavity is $\kappa/2\pi=250$ kHz. The transition frequency between the superconducting qubit is $\omega_{a}/2\pi=6.9$ GHz and the qubit relaxation rate is $\gamma/2\pi=1.8$ MHz, which indicates that the frequency detuning between the qubit and the coplanar cavity is $\Delta/2\pi=(\omega_{a}-\omega_{r})/2\pi=1.2$ GHz. The coupling strength between the qubit and the cavity is $g/2\pi=105$ MHz and the effective Stark shift induced per photon is $\chi=g^{2}/\Delta=2\pi\times9.188$ MHz, i.e., $\chi/\kappa\approx36.75$. In addition, the driving strengths ($\Omega_{R}$ and $\eta$) of the qubit and the cavity field are experimentally adjustable. In our simulations, we used the parameters $\gamma/\kappa=0.5$, $\chi/\kappa=15$, $\Omega_{R}/\chi=2$, and $\eta/\kappa=0.1$, which are experimentally accessible in circuit QED systems. These analyses indicate that the present scheme should be within the reach of current experimental techniques.

\section{Conclusion \label{conclusion}}

We have proposed an experimentally realizable method to recover the PB effect in a dispersively coupled atom-field system, in which the PB effect has been shown to vanish in the dispersive parameter regime. This is realized by introducing a transversal driving to the atom, and this model has been experimentally implemented by superconducting quantum circuits~\cite{Szombati2020,Schuster2007Resolving}. In the absence of the cavity-field driving, we have obtained the analytical eigenvalues and eigenstates of the driven dispersive JC model. We have also studied the PB effect by numerically calculating the photon-number distributions and the equal-time second-order correlation function of the cavity field. It was found that the PB effect can be observed in the single-photon resonant driving case. The influence of the system dissipation on the PB effect has been investigated by solving the quantum master equation. Our scheme presents a method to recover the PB effect in the strong dispersive JC model. Therefore, this work not only realizes the generation of the PB effect in a wider detuning parameter space of the JC model, but also provides inspiration for generating the conventional PB effect in dispersively coupled quantum systems.

\begin{acknowledgments}
J.-F.H. was supported in part by the National Natural Science Foundation of China (Grant No. 12075083), Scientific Research Fund of Hunan Provincial Education Department (Grant No. 18A007), and Natural Science Foundation of Hunan Province, China (Grant No. 2020JJ5345). J.-Q.L. was supported in part by National Natural Science Foundation of China (Grants No. 11774087, No. 11822501, No.~12175061, and No. 11935006), Hunan Science and Technology Plan Project (Grant No. 2017XK2018), and the Science and Technology Innovation Program of Hunan Province (Grants No. 2020RC4047 and No. 2021RC4029).
\end{acknowledgments}


\begin{thebibliography}{99}
\bibitem{footnote} We analyzed the energy spectrum of the dispersive JC model and checked the second-order correlation function of the cavity field in the steady state. The results indicate that there is no photon blockade effect in the dispersive JC model.
\end{thebibliography}
\end{document}